\documentclass[aps,pre,twocolumn,superscriptaddress,longbibliography,nofootinbib,showkeys,showpacs]{revtex4-2}

\usepackage[utf8]{inputenc}
\usepackage{amsfonts}
\usepackage{amsmath}
\usepackage{appendix}
\usepackage{bm}
\usepackage{braket}
\usepackage{caption}
\usepackage{color}
\usepackage{comment}
\usepackage{enumerate}
\usepackage[top=2cm,left=2cm,right=2cm,bottom=2cm]{geometry}
\usepackage{graphicx}
\usepackage{hyperref}
\usepackage[utf8]{inputenc}
\usepackage{mathptmx} 
\usepackage{scalerel}
\usepackage{subcaption}
\begin{document}
\title{Performance of quantum annealing for 2-SAT problems with multiple satisfying assignments}

\author{Vrinda Mehta\footnote{ORCID: 0000-0002-9123-7497}}
\affiliation{Institute for Advanced Simulation, J\"ulich Supercomputing Centre,\\
Forschungszentrum J\"ulich, D-52425 J\"ulich, Germany}
\author{Hans De Raedt\footnote{ORCID: 0000-0001-8461-4015}}
\affiliation{Institute for Advanced Simulation, J\"ulich Supercomputing Centre,\\
Forschungszentrum J\"ulich, D-52425 J\"ulich, Germany}
\author{Kristel Michielsen\footnote{ORCID: 0000-0003-1444-4262}}
\affiliation{Institute for Advanced Simulation, J\"ulich Supercomputing Centre,\\
Forschungszentrum J\"ulich, D-52425 J\"ulich, Germany}
\affiliation{RWTH Aachen University, 52056 Aachen, Germany}
\affiliation{AIDAS, 52425 J\"ulich, Germany}
\author{Fengping Jin\footnote{ORCID: 0000-0003-3476-524X}}
\affiliation{Institute for Advanced Simulation, J\"ulich Supercomputing Centre,\\
Forschungszentrum J\"ulich, D-52425 J\"ulich, Germany}
\date{\today}

\begin{abstract}
    Using a specially constructed set of hard 2-SAT problems with four satisfying assignments, we study the scaling and sampling performance of numerical simulation of quantum annealing as well as that of the physical quantum annealers offered by D-Wave. To this end, we use both the standard quantum annealing and reverse annealing protocols in both our simulations and on the D-Wave quantum annealer. In the case of ideal quantum annealing the sampling behavior can be explained by perturbation theory and the scaling behavior of the time to solution depends on the scaling behavior of the minimum energy gap between the ground state and the first excited state of the annealing Hamiltonian. The corresponding results from the D-Wave quantum annealers do not fit to this ideal picture, but suggest that the scaling of the time to solution from the quantum annealers matches those calculated from the equilibrium probability distribution. 
    
\end{abstract}

\maketitle

\section{Introduction}
Optimization problems are an important class of computational problems that find their applications across many real-world problems that include finance, medicine, logistics, scheduling, chemistry allocation of resources, etc. \cite{Orus2019,Woerner2019,venturelli2019reverse,sakuma2020application,solenov2018potential,cordier2022biology,wang2012design,tran2016hybrid,Mohammadbagherpoor:2021zqa,inoue2021traffic,Feld_2019,Micheletti_21,Bova2021}. All real-world optimization problems involve a large number of variables, which makes them intractable for exact solvers. This is mainly due to the commonly-observed exponential growth of the search-space as the size of such a problem grows, thus imposing a limitation on the computational resources required to implement a brute-force search for the optimal solution.

To circumvent this issue, various heuristic methods have been proposed for finding the solutions to the optimization problems, for example, gradient-based methods, variational methods, and simulated annealing \cite{nocedal2006numerical,fletcher2000practical,trefethen2022numerical,bertsekas1997nonlinear,press2007numerical,aarts1988simulated,kirkpatrick1983optimization,talbi2009metaheuristics,blum2003metaheuristics}. Quantum annealing is a metaheuristic algorithm, inspired by simulated annealing, wherein thermal fluctuations are replaced by quantum fluctuations \cite{apolloni1988numerical,finnila1994quantum,kadowaki1998quantum,farhi2000quantum,farhi2001quantum}. It is conjectured that owing to mechanisms like quantum tunneling, quantum annealing might be more efficient for searching the solution space of an optimization problem compared to simulated annealing where the state of the system can get trapped in a narrow potential barrier if it becomes high. The availability of commercial quantum annealers in recent years, for example, the ones offered by D-Wave Quantum Systems Inc. with more than 5000 qubits, has facilitated research in the direction of finding applications for quantum annealing and also in gauging its performance for solving them compared to other approaches \cite{Pudenz_2014,boixo2014evidence,ronnow2014defining,albash2018adiabatic,willsch2020benchmarking, calaza2021garden, willsch2022tailassign, montanez2023unbalanced, montanez2024unbalanced}.

Our previous studies \cite{paper1,paper2} have focused on the performance of the quantum annealing algorithm for solving optimization problems with unique solutions. In this paper, we explore the efficiency of the approach for solving problems that have more than one possible solution, which is the case for many real-world optimization problems. Apart from this, from a practical point of view, for real-world optimization problems, it might be useful to obtain solutions that satisfy all the constraints of the problem, even if they are not the optimal solutions. In such cases, the capability of quantum annealing to yield low-energy solutions, even if not the optimal ones, becomes a relevant measure for its performance.

The suitability of a certain method to solve problems with more than one solution can be judged using several criteria. The most obvious of them is the success probability, which is defined as the sum of the probabilities of obtaining the possible solutions for these problems. A better judge, and perhaps more standard of a metric is the scaling of the success probability or the time to solution (TTS) as a function of the problem size. Another relevant criterion for problems with more than one solution is the efficiency of the algorithm to yield all its solutions. This problem has been tackled in the past for the transverse Ising model in Ref.~\cite{matsuda2009ground}, where it was observed that for certain problems, the addition of higher-order transverse couplings to the annealing Hamiltonian might help alleviate the problem of unequal sampling probabilities of the ground states.

In this study, we focus on the sampling and scaling performance of the numerical implementation of the ideal standard quantum annealing algorithm as well as that of the ideal reverse annealing protocol and compare them with those obtained from a real quantum annealing system for solving sets of hard 2-satisfiability (SAT) problems with four satisfying assignments.

The standard algorithm for quantum annealing requires the system to start in the ground state of an easy-to-prepare initial Hamiltonian, typically chosen to be the uniform transverse field Hamiltonian, i.e.,
\begin{equation}
    H_I = -\sum_{i=1}^N \sigma_i^x,
\end{equation}
where $\sigma_i^x$ is a Pauli matrix. With the help of the annealing parameter, defined as $s=t/T_A$ where $T_A$ is the total annealing time, the system is slowly swept towards the problem Hamiltonian $H_P$ encoding the optimization problem to be solved, so that
\begin{equation}
     H(s) = A(s) H_I + B(s) H_P,
    \label{eq:annealing}
\end{equation}
where functions $A(s)$ and $B(s)$ controlling the annealing scheme are chosen such that $A(0)/B(0) \gg 1$ and $A(1)/B(1) \ll 1$. The problem Hamiltonian is the Ising Hamiltonian of the form
\begin{equation}
    H_P = -\sum_{i=1}^N h_i^z \sigma_i^z - \sum_{\langle i,j \rangle} J_{i,j}^z \sigma_i^z \sigma_j^z,
    \label{eq:hp}
\end{equation}
where $\sigma_i^z$ is a Pauli matrix, $h_i^z$ is the applied field acting along the $z$- direction, $J_{i,j}^z$ is the coupling between the $i$th and $j$th spins, and $\langle i,j \rangle$ denotes the set of coupled spins.

A comparatively less explored variation of the quantum annealing algorithm is the reverse annealing protocol. Starting from one of the low-lying eigenstates of Eq.~(\ref{eq:hp}), the system is slowly swept backward (by decreasing $s$ and therefore increasing the strength of the transverse field) till a certain reversal distance $s_r$ \cite{perdomo2011study,chancellor2017modernizing,king2018observation,ottaviani2018low,venturelli2019reverse,rieffel,reverse_lidar,dynamics_lidar}. From there, the system continues again towards $s=1$, like in the standard annealing algorithm, with an optional waiting time $t_{wait}$ at $s_r$. It is conjectured that doing this might be able to yield a better solution to the encoded optimization problem than the state in which the algorithm started \cite{reverse_lidar}. In this work, we explore this version of annealing to study its efficiency in the context of fairly sampling all the degenerate ground states of the problem Hamiltonian using both simulations and D-Wave Advantage\_5.1 (DWAdv).

Our results show that the sampling behavior of the ideal standard quantum annealing (in the long annealing time regime and in the absence of temperature effects and noise) can be explained using perturbation theory. This is in agreement with the idea proposed in Ref.~\cite{fairsampling_troyer}. This, however, is not found to be the case for sampling probabilities obtained using DWAdv, which are almost uniform for most of the problems studied. On the other hand, shifting our focus from the ensemble of the 2-SAT problems to a specific 14-variable instance that is found to have an almost zero sampling probability of one of the ground states using the standard quantum annealing simulations, we find that the sampling behavior obtained using the simulations for an ideal implementation of reverse annealing vastly depends on the choice of the relevant parameters. Furthermore, in this case, we note a better match of the sampling probabilities from DWAdv with those obtained numerically.

Furthermore, using both the numerical implementation of standard quantum annealing and DWAdv, we find an exponential scaling of the time to solution as a function of the size of the problem, as was the case for the 2-SAT problems with a unique satisfying assignment in Ref.~\cite{paper2}, although the scaling exponents obtained with DWAdv are significantly smaller.

The content of this paper has been organized as follows. In Sec.~\ref{sec:probs}, we discuss first the problem sets used for this work. Section~\ref{sec:standard} showcases the sampling and scaling results using standard quantum annealing for
degenerate problem Hamiltonians. In Sec.~\ref{sec:samp_rev}, we show the results for the sampling behavior of the degenerate ground states using reverse annealing obtained from simulations and DWAdv. Lastly, we summarize our observations in Sec.~\ref{sec:conclusion}.

\section{Problem sets}
\label{sec:probs}
In this work, we consider sets of hard 2-SAT problems for testing the scaling and the sampling efficiency of quantum annealing. A 2-SAT problem is made up of several clauses, each consisting of two Boolean literals (a Boolean variable $x_i$ or its negation $\overline{x_i}$ for $i=1,...,N$), i.e.,

\begin{equation}
    F = (L_{1,1} \lor L_{1,2}) \land (L_{2,1} \lor L_{2,2}) \land ... \land (L_{M,1} \lor L_{M,2}),
    \label{eq:2sat}
\end{equation}
where $L_{\alpha,j}$ represents the $j$th literal in the $\alpha$th clause for $j = 1,2$ and $\alpha=1,..,M$. A solution to the 2-SAT problem is then an assignment to the variables $x_i$'s that make each clause true, and hence the 2-SAT problem satisfiable. 

In the following sections, we first describe the employed method for creating these problems, and subsequently some of the properties of the resulting sets of problems. We then discuss the mapping of these problems to a form suitable for quantum annealing for solving them.

\subsection{Creation of the problems}
As the first step in creation of 2-SAT problems, we fix the number of clauses $M$. Since the satisfiability threshold, defined as the ratio of the number of clauses to that of the variables ($N$) for which a SAT problem changes from being satisfiable in the mean to being unsatisfiable in the mean, lies around $M \approx N$ for 2-SAT problems, we choose $M=N+1$. Each clause is then made to satisfy the following constraints :
\begin{itemize}
    \item the two literals chosen for a clause should correspond to different variables, 
    \item each variable should be used at least once in one of the clauses, 
    \item none of the clauses should be repeated.
\end{itemize}

After obtaining the 2-SAT problems with clauses subject to the above-mentioned constraints, we first identify the problems which are satisfiable. For this, we use the Kosaraju-Sharir's algorithm \cite{sharir1981strong} which identifies the strongly connected components (sets of vertices reachable from one another) for every problem from its implication graph, and if a variable and its negation are found to belong to the same strongly connected component, the given 2-SAT problem is unsatisfiable. Next, we discard the problems which are not satisfiable, and from the resulting set of problems we find the number of satisfying assignments of every problem using brute-force search. We then select the problems with 1, 2, or 4 satisfying assignments. Since the number of possible assignments grows exponentially with the size of the problems, it is not possible to obtain problems with a large number of variables in this way. For this reason, we restrict ourselves to problem sets with $6 \leq N \leq 20$. To this end, we used workstations equipped with Intel Core i7-8700 and 32 GB memory for problems with $N \leq 13$, while for the larger problems we employed the supercomputer JUWELS of the J\"{u}lich Supercomputing Centre at Forschungszentrum J\"{u}lich \cite{JUWELS}. Each set corresponding to a given $N$ and one of the chosen values for the number of satisfying assignments has at least 100 problems.

\subsection{Properties of the 2-SAT problems}
Next, we focus on discussing some of the properties of the sets of problems that have been obtained as explained in the previous section. 

For a K-SAT problem, the average number of satisfying assignments $\mu$, obtained using combinatorics, is given by
\begin{equation}
    \langle \mu \rangle = \left ( 1 - \frac{1}{2^K} \right ) ^ M 2^N, \\
    \label{eq:deg_formula}
\end{equation}
when the way in which the clauses are made is not subject to any constraints.

\begin{figure}
    \centering
    \includegraphics[scale=0.7]{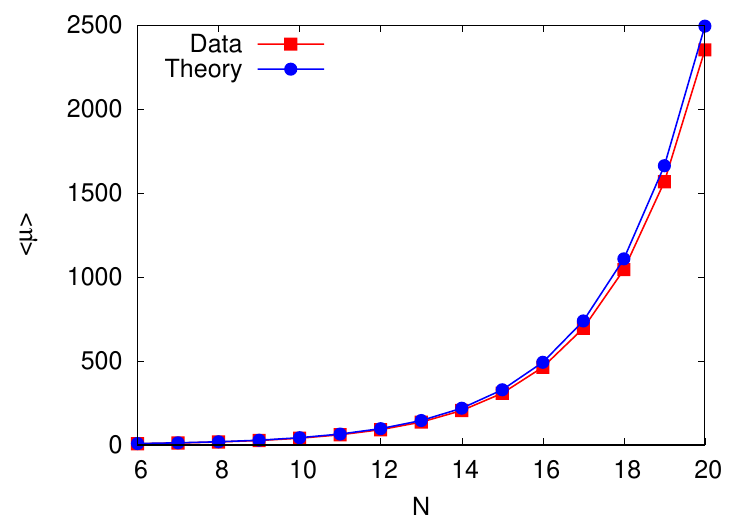}
    \caption{(Color online) Average number of satisfying assignments $\mu$ for the 2-SAT problems as a function of the problem size where $6 \leq N \leq 20$ and $M = N+1$.}
    \label{fig:DegVsN}
\end{figure}

In Fig.~\ref{fig:DegVsN} we show the average degeneracy of the ground state of the 2-SAT problems, that is the average number of satisfying assignments, as a function of the problem size for $M = N+1$ for $6 \leq N \leq 20$.
From this figure, it is clear that for the obtained set of 2-SAT problems, the average degeneracy of the ground state matches well with its theoretical estimate according to Eq.~(\ref{eq:deg_formula}), although there are some differences in the two values. These are a consequence of the additional constraints that are imposed while creating the clauses of the 2-SAT problems.

\begin{figure}
    \centering
    \includegraphics[scale=0.7]{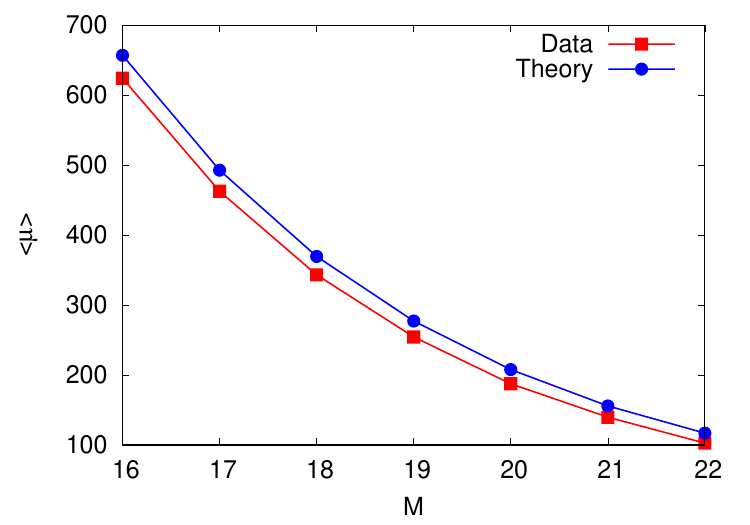}
    \caption{(Color online) Average number of satisfying assignments $\mu$ for the 2-SAT problems as a function of the number of clauses of the problem size where $M=N+1$ for problems with 1, 2, and 4 satisfying assignments.}
    \label{fig:DegVsM}
\end{figure}

Next, while keeping the problem size fixed at $N=16$, we study the dependence of the average degeneracy of the ground state of the 2-SAT problems as a function of the number of clauses $M=N+c$, where $c$ varies from 0 to 6. The corresponding result is shown in Fig.~\ref{fig:DegVsM}. While there is an overall similarity in the trend of the average degeneracy, also here there are slight deviations between the theoretical value of the degeneracy and the one obtained for the sets of 2-SAT problems created which can be attributed to the way in which these problems are created. Additionally, it is evident from Fig.~\ref{fig:DegVsN} and Fig.~\ref{fig:DegVsM} that it becomes progressively difficult to find 
problems with ground state degeneracy 1, 2, or 4. This limits the number of problems that can be created in this way.

\begin{figure}
    \centering
    \includegraphics[scale=0.7]{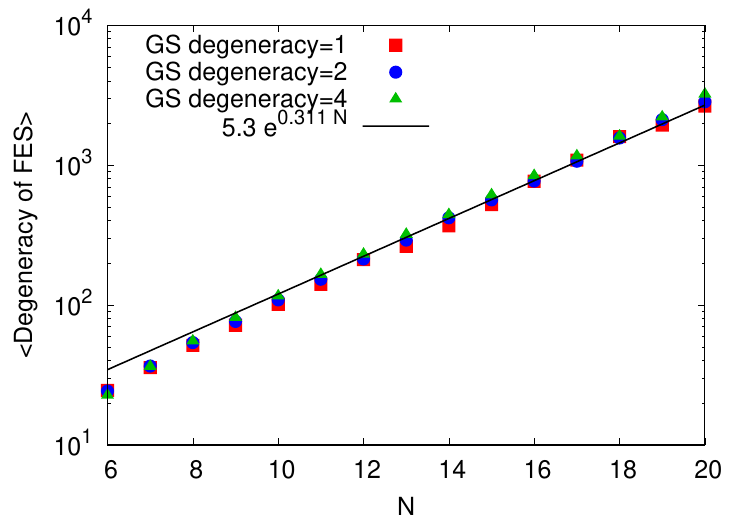}
    \caption{(Color online) Average degeneracy of the first excited states (FES) of the problem Hamiltonians corresponding to the 2-SAT problems as a function of the problem size where $6 \leq N \leq 20$ and $M = N+1$.}
    \label{fig:FESDeg}
\end{figure}

We now look at the scaling of the average degeneracy of the first excited state (FES) for the obtained sets of 2-SAT problems with 1, 2, and 4 satisfying assignments as a function of the problem size, where $6 \leq N \leq 20$ and $M = N+1$. This is shown in Fig.~\ref{fig:FESDeg}. From these results, we see that for all three cases with different ground state degeneracies, the average degeneracy of the first excited state increases exponentially as the problem size grows with a similar scaling exponent of 0.311. This suggests that for every two additional variables in these problems the degeneracy of the first excited state almost doubles.

\subsection{Reformulation of 2-SAT problems as Ising Hamiltonian}

To employ quantum annealing for solving the 2-SAT problems, we first need to bring them to a form that is suitable for the quantum annealing algorithm. For the D-Wave systems, this is the QUBO or Ising representation of the problem. In this work, we choose to reformulate the obtained set of problems to Ising problems. For this, we map every clause of a given 2-SAT problem to the Ising model with Hamiltonian
\begin{equation}
        C_{2SAT} = \sum_{\alpha=1}^M (\epsilon_{(\alpha,1)} s_{i[\alpha,1]}-1)(\epsilon_{(\alpha,2)} s_{i[\alpha,2]}-1),
        \label{eq:map2SAT}
    \end{equation}
where $i[\alpha,j]$ represents the variable $i$ that is involved in the $j$th term of the $\alpha$th clause for $i=1,\dots,N$, $j=1,2$, and $\alpha=1,\dots,M$. If this variable is $x_i$ then $\epsilon_{(\alpha,j)}=1$, whereas if it is its negation $\overline{x_i}$ then $\epsilon_{(\alpha,j)}=-1$.

At this point, it should be noted that while 2-SAT problems are not NP-hard problems as decision problems, the problem of finding all the solutions of a 2-SAT problem is.
\begin{comment}
and the best classical solvers can solve 2-SAT problems in polynomial time, for quantum annealing the Ising representation of a 2-SAT problem is indistinguishable from that of an NP-hard spin-glass problem. However, the 2-local terms that naturally arise from the Ising mapping of the 2-SAT problems make these problems to be more suitable for physical quantum annealers like those from D-Wave.
\end{comment}

\section{Standard quantum annealing}
\label{sec:standard}

As discussed previously, the standard quantum annealing Hamiltonian starts from the transverse Ising Hamiltonian, with a decreasing strength of the transverse field and an increasing strength of the longitudinal fields and couplings. For numerically implementing the dynamics of quantum annealing, we make use of the second-order product formula algorithm \cite{paper1,paper2}. The sampling probabilities of the four ground states are then obtained by computing the overlap of the resulting state with the known ground states of the problem Hamiltonian. On the other hand, the sampling probabilities using the quantum annealer are obtained by determining the ratio of the number of times one of the ground states is sampled to the total number of samples. Using the above-mentioned set of 2-SAT problems with four satisfying assignments, we first assess the fairness of quantum annealing in sampling the four ground states of the problem Hamiltonian. Furthermore, using the obtained success probabilities we also study the scaling of the time to solution (TTS) and inverse of success probability ($1/p$) obtained using quantum annealing for solving these problems and compare it with that for the 2-SAT problems with a unique solution studied previously \cite{paper2}.

\subsection{Sampling efficiency}
\label{sec:Sampling_standard}

We start our analysis by focusing on the efficiency of standard quantum annealing to fairly sample the four ground states of the 2-SAT problems fairly. More specifically, we test if quantum annealing can yield the different degenerate ground states of a problem with comparable probabilities. %For this, we make use of both simulations as well as DWAdv, and study the sampling probabilities of the ground states of the given problems. The observations from the ideal implementation of the algorithm in the long annealing time limit are then explained using perturbation theory.

\subsubsection{Simulation results}
\label{sec:std_samp_sim}

We start by discussing the numerically obtained sampling results for solving the given set of 2-SAT problems with four satisfying assignments using standard quantum annealing. To this end, we choose three different annealing times, namely, $T_A = 10, 100, 1000$, and record the resulting sampling probabilities for all the 100 problems belonging to sets with $6 \leq N \leq 20$. It is worth noting here that in the D-Wave energy scales, the annealing times $T_A=10$, $T_A=100$, and $T_A=1000$ in our simulations approximately correspond to 0.5~ns, 5~ns, and 50~ns. 

\begin{figure}
    \centering
    \includegraphics[scale=0.7]{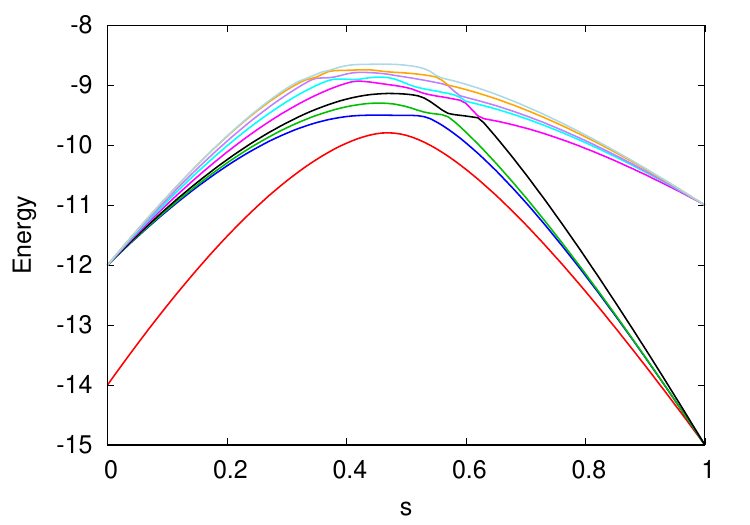}
    \caption{(Color online) Energy spectrum of a 14-variable 2-SAT problem Hamiltonian labeled as the problem "230" with four degenerate ground states.}
    \label{fig:fairsampling_spec}
\end{figure}

The first observation is that for a majority of the problems, the sampling probabilities of the four ground states are comparable. %when the annealing time is sufficiently long, i.e., for $T_A=100$ and $T_A=1000$. On the other hand, while for the smaller problem sizes, we observe a similar trend of the sampling probabilities with $T_A=10$, as the size of the problems increases the sampling behavior is \textcolor{blue}{not necessarily} fair\textcolor{blue}{, by which we mean that the resulting sampling probabilities might not be comparable}.
This observation can be understood based on the typical energy spectra of these problems, an example of which is shown in Fig.~\ref{fig:fairsampling_spec} for a 14-variable 2-SAT problem. We find that the energy spectrum exhibits a concatenation of anticrossings between the ground state of the annealing Hamiltonian and its fourth excited state, i.e., the anticrossings between the subsequent energy levels occur at increasing values of the annealing parameter $s$. When the annealing times are not sufficiently long for an adiabatic evolution, this arrangement facilitates the leakage of the amplitude present in the ground state out of the ground state subspace. In such cases, quantum annealing might fail in finding all the solutions to the problem.

\begin{comment}
Since the time to solution is a function of the success probability of the algorithm for a given annealing time, the exponential increase in TTS99 in Fig.~\ref{fig:tts99_deg_simulations} suggests an exponentially decreasing total success probability as the size of the problem grows. Therefore, while the annealing time $T_A=10$ suffices to yield a reasonable total success probability for the smaller problems, the total success probability rapidly decreases as the size of the problems is increased. Similar observations have also been noted in our previous work for problems with a unique solution \cite{paper1,paper2}. This has a direct impact on the sampling probabilities of the degenerate ground states of the problems considered here.
\end{comment}

Inspecting more closely, we find that while for a majority of the problems under consideration, the sampling probabilities of the four ground states are fair for long annealing times, there are a considerable number of problems for which it is not the case. In what follows, we focus on three examples of 14-variable 2-SAT problems, specified in Table~\ref{tab:SATproblems} of Appendix~\ref{sec:appendix_problems}, that exhibit three distinct sampling behaviors that are observed in the regime of sufficiently long annealing times. 

\begin{itemize}
    \item Fair sampling: As our first case we choose the 14-variable problem labeled as problem "1" for which quantum annealing can yield the four ground states fairly for $T_A = 100$, and $T_A = 1000$. However, for the short annealing time $T_A=10$, the state of the system can deviate from the ground state following the cascade of anticrossings that these problems typically exhibit in the spectrum (see Fig.~\ref{fig:fairsampling_spec}). This can lead to a significant decrease in the total success probability and the sampling behavior of the four ground states might vastly vary.

    \item Unfair sampling: For a few other cases like problem labeled "3" in Table~\ref{tab:SATproblems}, the sampling probabilities of the four ground states remain unequal even for long annealing times. Table~\ref{tab:fa_prob_3} shows the sampling probabilities for this problem for different annealing times, from where it is clear that the sampling probabilities of the ground states $\ket{\psi_0^1}$ and $\ket{\psi_0^4}$ remain significantly different from those of the states $\ket{\psi_0^2}$ and $\ket{\psi_0^3}$ for all the chosen annealing times.
    \begin{table}[!htp]
    \begin{center}
    \caption{Sampling probabilities of the four degenerate ground states $\ket{\psi_0^i}$, $i=1,2,3,4$, of problem "3", corresponding to different annealing times $T_A$, as obtained by standard quantum annealing.}
    \begin{tabular}{ |c|c|c|c| }
     \hline
     \textbf{State} & \bm{$T_A=10$} & \bm{$T_A=100$} & \bm{$T_A=1000$}\\ 
     \hline
     $\ket{\psi_0^1}$ & 0.0219 & 0.0917 & 0.1388 \\ 
     \hline
     $\ket{\psi_0^2}$ & 0.0622 & 0.2321 & 0.3612\\ 
     \hline
     $\ket{\psi_0^3}$ & 0.0622 & 0.2321 & 0.3612\\ 
     \hline
     $\ket{\psi_0^4}$ & 0.0219 & 0.0917 & 0.1388\\ 
     \hline
     Total & 0.1682 & 0.6476 & 1.000\\ 
     \hline
    \end{tabular}
    \label{tab:fa_prob_3}
    \end{center}
    \end{table}

    \item Total suppression: In more extreme cases, we find that the sampling probability of one of the ground states can be totally suppressed. One such 14-variable problem is given in Table~\ref{tab:SATproblems} referred to as problem "230". The sampling probabilities of the ground states for this problem are given in Table~\ref{tab:fa_prob_230}, from where it can be seen that the sampling probability of the ground state $\ket{\psi_0^3}$ is nearly zero for $T_A=1000$. Such problem instances are therefore interesting to study further, for example, using reverse annealing.

    \begin{table}[!htp]
    \begin{center}
    \caption{Sampling probabilities of the degenerate ground states $\ket{\psi_0^i}$, $i=1,2,3,4$, of problem "230", corresponding to different annealing times $T_A$, as obtained by standard quantum annealing Hamiltonian.}
    \begin{tabular}{ |c|c|c|c| }
     \hline
     \textbf{State} & \bm{$T_A=10$} & \bm{$T_A=100$} & \bm{$T_A=1000$}\\ 
     \hline
     $\ket{\psi_0^1}$ & 0.1233 & 0.4427 & 0.4986 \\ 
     \hline
     $\ket{\psi_0^2}$ & 0.0742 & 0.2605 & 0.2507\\ 
     \hline
     $\ket{\psi_0^3}$ & 0.0648 & 0.0131 & 9.56$\times 10^{-10}$ \\ 
     \hline
     $\ket{\psi_0^4}$ & 0.0589 & 0.2214 & 0.2506\\ 
     \hline
     Total & 0.3212 & 0.9377 & 0.9909\\ 
     \hline
    \end{tabular}
    \label{tab:fa_prob_230}
    \end{center}
    \end{table}

\end{itemize}

\subsubsection{D-Wave results}
After having seen the behavior of ideal quantum annealing using the standard annealing Hamiltonian for sampling problems with degenerate ground states, we perform similar experiments using DWAdv. We choose here the default value $T_A=20~\mu s$ for the annealing time and set the number of samples to 1000. In this case we find comparable sampling probabilities of the four ground states for nearly all the problems in the set. Even for the problem "230" which was noted to have a totally suppressed sampling probability of the ground state $\ket{\psi_0^3}$ in our simulations for $T_A=1000$, the four sampling probabilities from DWAdv are given as 0.1552, 0.2470, 0.2765, and 0.3124. Since the annealing time chosen for our runs on DWAdv is much longer than those used in the simulations, the differences in the two sampling behaviors are indicative of the presence of noise and temperature effects in the D-Wave system\, which in this case, can be beneficial for finding all the solutions to the given problems.

\subsubsection{Perturbation theory}
In section~\ref{sec:std_samp_sim}, we discussed the three kinds of sampling behaviors that were observed for our set of 2-SAT problems with four satisfying assignments using quantum annealing. In this section, we attempt to understand the reasons behind this using perturbation theory. In the long annealing time limit, the sampling behavior of the ideal quantum annealing algorithm can be determined by the overlap of the ground states of the problem Hamiltonian with the ground state of the instantaneous Hamiltonian in the vicinity of $s \approx 1$. The instantaneous ground state of this Hamiltonian can be obtained using perturbation theory by treating the initial Hamiltonian $H_I$ as a small perturbation to the problem Hamiltonian $H_P$. However, since the ground state of the problem Hamiltonian is degenerate, the choice for the basis vectors for the degenerate subspace of the problem Hamiltonian becomes arbitrary (since in this subspace the problem Hamiltonian is equivalent to the identity matrix times the ground state energy). In order to keep perturbation theory going, one needs to choose an appropriate basis, i.e., a basis that diagonalizes the perturbation matrix in the degenerate subspace. The eigenvectors of perturbation matrix $V = \braket{\psi_0^i|H_I|\psi_0^j}$ are thus a valid choice for the basis, where $\ket{\psi_0^i}$ are the ground states of the problem Hamiltonian in the computational basis and for $i,j=1,2,3,4$. If the lowest eigenvalue of the perturbation matrix $V$ is non-degenerate, the addition of perturbation to the problem Hamiltonian lifts the degeneracy of the latter. The sampling probabilities of the ground states of the problem Hamiltonian are then given as $|a_i|^2$ where $\nu_0 = \sum_{i=1}^4 a_i \ket{\psi_0^i}$ is the ground state of the perturbation matrix. Next, we take up the  example problems "1" and "230" presented in section~\ref{sec:Sampling_standard}, and use perturbation theory to calculate the theoretical sampling probabilities.

The four ground states of problem "1" are $\ket{\psi_0^1} =\ket{10101010000000}$, $\ket{\psi_0^2} =\ket{10101011000000}$, $\ket{\psi_0^3} =\ket{10101010000100}$, and $\ket{\psi_0^4} =\ket{10101011000100}$. The first-order perturbation matrix for this problem is given by
\begin{equation}
    V =
    \begin{pmatrix}
         0 & -1 & -1 & 0 \\
        -1 & 0 & 0 & -1 \\
        -1 & 0 & 0 & -1 \\
         0 & -1 & -1 & 0 \\
    \end{pmatrix}.
\end{equation}
The ground state of this perturbation matrix is $\ket{\nu_1} = 1/2(1,1,1,1)$. The sampling probabilities of the four ground states of this problem can therefore be expected to be 0.25, as is found to be the case for our simulations with long annealing times. On the other hand, for problem "230" we have $\ket{\psi_0^1} =\ket{11110100100101}$, $\ket{\psi_0^2} =\ket{11110100110101}$, $\ket{\psi_0^3} =\ket{10100101000111}$, and $\ket{\psi_0^4} =\ket{11110100100111}$.
The perturbation matrix for this problem is thus
\begin{equation}
    V =
    \begin{pmatrix}
        0 & -1 & 0 & -1 \\
        -1 & 0 & 0 & 0 \\
        0 & 0 & 0 & 0 \\
        -1 & 0 & 0 & 0 \\
    \end{pmatrix}.
    \label{eq:matrix_suppressed}
\end{equation}

It is evident from Eq.~(\ref{eq:matrix_suppressed}) that the ground state $\ket{\psi_0^3}$ of the problem is decoupled from the rest of the subspace. The ground state of this matrix is given by $\ket{\nu_1} = 1/2 (\sqrt{2},1,0,1)$, and thus for sufficiently long annealing times the sampling probabilities of the four ground states can be expected to be 0.5, 0.25, 0, and 0.25, respectively. These values are in close agreement with the sampling probabilities shown in Table~\ref{tab:fa_prob_230} obtained numerically for this problem for $T_A=1000$. On the other hand, in cases where non-adiabatic mechanisms play a significant role in the evolution of the system, for example in our simulations corresponding to $T_A=10$ or for systems where temperature effects and noise are present, the state of the system can leak out of the ground state subspace, and the ground state of the instantaneous Hamiltonian no longer dictates the sampling behavior.

As discussed above, when sampling all solutions is necessary, the standard quantum annealing Hamiltonian—featuring only single $\sigma^x$
  terms in the initial Hamiltonian—may be insufficient. To achieve a more uniform sampling of ground states, one possible approach, as noted in Refs. \cite{Nishimori2008,fairsampling_troyer,zhangGrover2025}, is to incorporate higher-order transverse couplings into the initial Hamiltonian.

\subsection{Scaling performance}
\label{sec:scaling}

So far, we looked at quantum annealing using the standard annealing Hamiltonian only in the context of its efficiency in sampling the multiple ground states of problems with more than one solution. We now shift our focus towards the scaling of TTS99 using quantum annealing for solving such problems. TTS99 is defined as the compute time required to obtain the solution to the optimization problem at least once in multiple anneals, with 99\% certainty. Mathematically,
\begin{equation}
    TTS = \frac{\ln{(1-P_{target})}}{\ln({1-p})} T_A,
    \label{eq:TTS}
\end{equation}
where $P_{target}$ is the target probability, and $p$ is the total success probability obtained from a single run of the algorithm with an annealing time $T_A$, which is the sum of the sampling probabilities of the four ground states. In the following, we discuss the scaling performance of standard quantum annealing using the results obtained from both simulations and the D-Wave quantum annealer DWAdv.

\subsubsection{Simulation results}
\label{sec:scaling_sim}
Using the numerically obtained total success probabilities for the 2-SAT problems with $6 \leq N \leq 20$ and four solutions, we study the scaling of the average TTS99 of these problems for the three annealing times $T_A=10,100,1000$.

\begin{figure}
\centering
\includegraphics[scale=0.7]{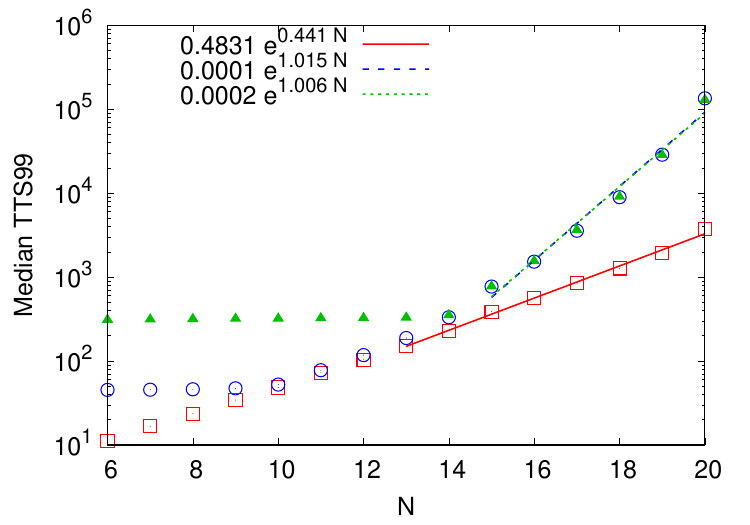}
     \caption{(Color online) Numerically obtained scaling of median TTS99 as a function of the number of variables for $T_A=10$ (square), $T_A=100$ (circle), and $T_A=1000$ (triangle).}
\label{fig:tts99_deg_simulations}
\end{figure}

Figure~\ref{fig:tts99_deg_simulations} shows the scaling of the median TTS99 as a function of the system size $N$ for annealing times $T_A=10,100,1000$. As expected from our previous scaling results for 2-SAT problems with a unique solution studied in Ref.~\cite{paper2}, TTS99 is also found to be exponentially growing with $N$ for the 2-SAT problems with four satisfying assignments.
The median TTS99 in this case scales with exponents $r_{TTS99}=0.441$, 1.015, and 1.006 for $T_A=10$, 100, 1000, respectively. Although these values are slightly better compared to the scaling exponents for the non-degenerate problem Hamiltonians obtained using the standard quantum annealing Hamiltonian from \cite{paper2}, in the long annealing time limit, the scaling behavior of the algorithm is still worse compared to a simple enumeration of all the possible assignments which scales with an exponent of 0.693. %However, for short annealing times like $T_A=10$, the scaling exponent is significantly smaller. This observation was also noted for the non-degenerate problem Hamiltonians \cite{paper2}, and was explained on the basis of the non-adiabatic mechanism of fast annealing.

Furthermore, from Fig.~\ref{fig:tts99_deg_simulations} we observe that while for $T_A=1000$ the median TTS99 values remain constant as the size of the problems initially increases, they increase exponentially for $N \geq 14$, and the median values coincide with those corresponding to $T_A=100$. Similarly, $T_A=100$ also results in constant values of median TTS99 for a smaller initial range of $N$ compared to $T_A=1000$, before increasing exponentially with the problem size and coinciding with those corresponding to $T_A=10$ for a few intermediate values of $N$. One can understand the reasons behind these observations from the transition probability versus annealing time scan. This aspect is discussed in Appendix~\ref{app:trans_prob}.

%This argument holds as long as the chosen annealing time is sufficiently long for the system to still be described as a two-level system with transitions occurring only close to the anticrossing. 

%However, unlike those problems, the number of problems that have a larger success probability for $T_A=10$ than the longer annealing times is much smaller for the problems with four satisfying assignments. This can be seen from Fig.~\ref{fig:deg_successs_comparison_T10} showing the success probability corresponding to $T_A=100$ and $T_A=1000$ against that corresponding to $T_A=10$. This can once again be understood on the basis of the energy spectrum of the annealing Hamiltonian shown in Fig.~\ref{fig:fairsampling_spec}. Owing to the presence of cascade of anticrossings between the ground state and the higher excited states, the probability of the amplitude transferred to the higher excited states (before the anticrossing between the instantaneous ground state and the first excited state) to return to the ground state (at the anticrossing) is lower. 

\subsubsection{D-Wave results}
\label{sec:scaling_dwave}
\begin{figure}
\centering
\includegraphics[scale=0.7]{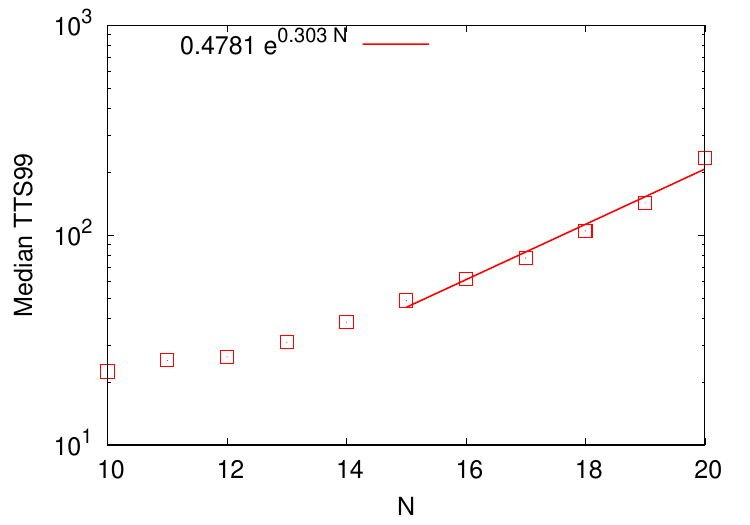}
\caption{(Color online) Scaling of median TTS99 as a function of the number of variables obtained using DWAdv for $T_A=20~\mu s$.}
\label{fig:tts99_deg_dwave}
\end{figure}

Focusing next on the results for the success probability from the DWAdv system for solving the sets of problems with degenerate problem Hamiltonians, Fig.~\ref{fig:tts99_deg_dwave} shows the scaling of the median TTS99 for $T_A=20~\mu s$. As was the case for the non-degenerate problems in Ref.~\cite{paper2}, the scaling exponent for the given problem sets is also significantly smaller using the D-Wave system than that obtained from simulations. In this case, the median TTS99 is found to scale with an exponent of 0.303, which is also significantly smaller than the brute force search exponent of 0.693. This observation suggests that noise and temperature effects are dominant in the system and can be, in some cases, advantageous for the performance of quantum annealing.

\begin{figure}
\centering
\includegraphics[scale=0.7]{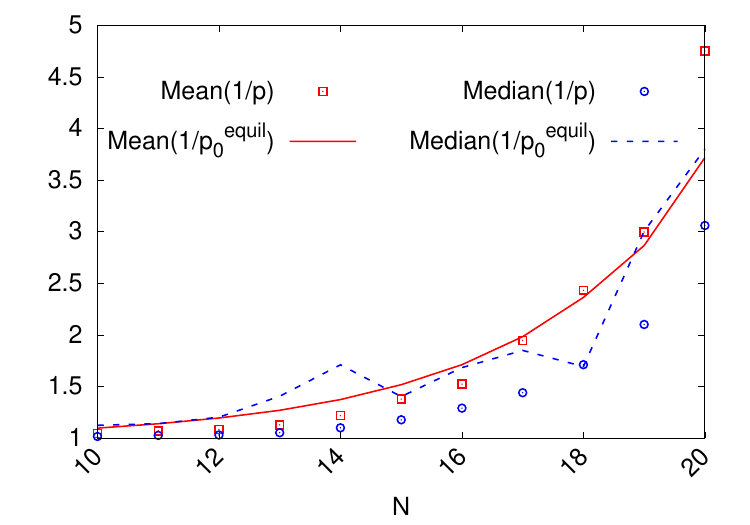}
\caption{(Color online) Comparison of scaling of the mean and the median cases obtained from DWAdv with those of the equilibrium 1/p for $T_A=20$~$\mu s$.}
\label{fig:alpha_combined}
\end{figure}

Since our results indicate that DWAdv is not an ideal quantum annealer, as an extension to our analysis, we also compare the obtained scaling performance to that of the equilibrium probability distribution. From the D-Wave results, we mainly note the contribution of only the lowest two energy levels, which is a manifestation of a low temperature, and therefore we restrict ourselves to these two levels. The equilibrium ground state probability for the ground state is then given by

\begin{equation}
    p_0^{equil} = \frac{1}{1+\frac{g_1}{g_0} e^{-\beta \Delta E}},
    \label{eq:equil}
\end{equation}
where $g_0 = 4$ is the ground state degeneracy, and $g_1$ is that of the first-excited state, $\beta = C/T$ for $C=h\, B(s=1)\times{10}^{9}/(2 k_B)=0.206$ and some corresponding temperature $T$ (expressed in kelvin), and $\Delta E$ is the energy between the lowest two energy levels of the 2-SAT problem. 
Figure~\ref{fig:alpha_combined} shows a plot of $\langle1/p\rangle$ obtained from DWAdv as a function of the problem size, in comparison to that of $1/p_0^{equil}$ obtained from Eq.~(\ref{eq:equil}) using the average value of the first excited state degeneracy for each problem set corresponding to an $N$ and $\beta$ as a fitting parameter. In this case, we obtain $\beta=1.42$ which yields $T\approx145$~mK. Furthermore, from each set of problems, we show the success probabilities of one of the problems constituting the median $Med(1/p)$ of the DWAdv results, and calculate the corresponding equilibrium success probabilities for these cases, with the fitting parameter $\beta=1.53$ which corresponds to $T\approx 135$~mK. From these results, we observe a good agreement between the results from the quantum annealer and the corresponding analytical expression, especially in the case of $\langle 1/p \rangle$. In appendix~\ref{app:otherprobs}, we test this idea for two other sets of problems and find the trend to still hold.

\section{Sampling efficiency of reverse annealing}
\label{sec:samp_rev}
We now shift our focus to the sampling efficiency of the reverse annealing protocol using both simulations and DWAdv. In contrast to the previous section, we divert our attention from the level of the ensemble of the 2-SAT problems to an interesting instance of a 14-variable, namely instance 230 whose third ground state is found to have a zero sampling probability using standard quantum annealing. Starting from one of the known ground states of this problem, we analyze the sampling behavior of the four ground states at the end of the protocol for different choices of relevant parameters. It is worth mentioning that one of the main motivations for using reverse annealing is to find states with lower energies as compared to the initial state, our analysis cannot improve the solution quality as we already start from the lowest energy solution. However, this choice for the initial state is reasonable for studying the sampling behavior of the protocol.

\subsubsection{Simulation results}
We start by addressing the numerically obtained results for an ideal implementation of reverse annealing, i.e., in the absence of temperature and noise effects. In what follows, we discuss the effects of controls like annealing time, reversal distance, waiting time, and the initial state on the sampling probabilities of the four ground states, one by one.

    \textbf{Different annealing times:} To study the effects of the annealing time on the sampling probabilities of the four ground states, we first fix the values of the other parameters. We choose $s_r = 0.7$, $T_W = 0$, and start with the ground state $\ket{\psi_0^1}$ as the initial state. It should be noted that while these choices might not be optimal or useful in cases where sampling other low-energy states is the motivation for using reverse annealing, they are reasonable choices if the aim is to sample other ground states.
    
\begin{figure*}
     \begin{minipage}{0.33\textwidth}
         \centering
         \includegraphics[width=\textwidth]{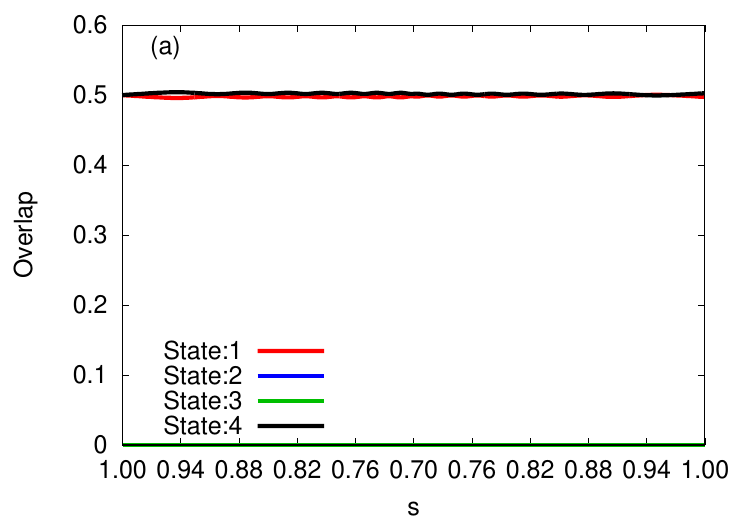}
     \end{minipage}
     \hfill
     \begin{minipage}{0.33\textwidth}
         \centering
         \includegraphics[width=\textwidth]{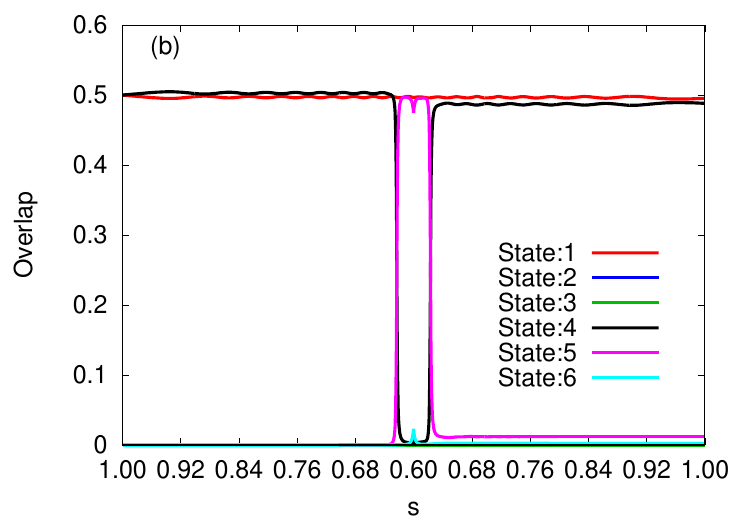}
     \end{minipage}
        \begin{minipage}{0.33\textwidth}
         \centering
         \includegraphics[width=\textwidth]{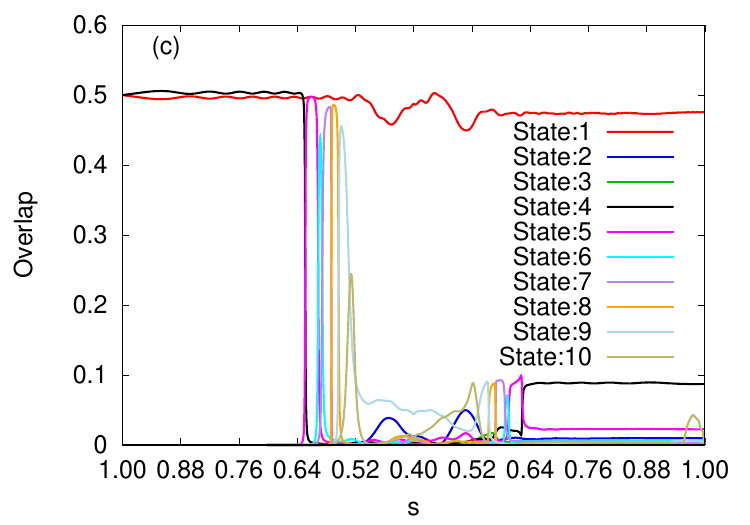}\\
     \end{minipage}
     \hfill
     \caption{(Color online) Overlap of the state of the system with the four lowest energy instantaneous eigenstates of the Hamiltonian for (a) $s_r=0.7$, (b) $s_r=0.6$, and (c) $s_r=0.4$ and $T_A=100$ during the reverse annealing segment ($s$ changes from $1$ to $s_r$) followed by the annealing segment ($s$ changes from $s_r$ to $1$).}
    \label{fig:reverse_overlap}
\end{figure*}

    In Table~\ref{tab:prob_ta_state1}, we show the sampling probabilities of the four ground states of problem "230" for various annealing times, and various observations can be made. Firstly, we see that the total success probability is nearly one for all the annealing times, especially for the longer ones. This can be understood in connection with Fig.~\ref{fig:fairsampling_spec}, which shows that in the regime of $s \geq 0.7$, there are no anticrossings between the lowest four ground states and the higher excited states which could result in a leakage of the state of the system from the ground state subspace. The second observation worth noting is that the sampling probability of the ground state $\ket{\psi_0^3}$ remains low for all annealing times, and decreases as the annealing time increases. As before, this can be explained on the basis of Eq.~(\ref{eq:matrix_suppressed}), according to which the ground state $\ket{\psi_0^3}$ of the problem Hamiltonian, or equivalently, the second excited state of the instantaneous Hamiltonian, remains decoupled from the other three ground states. Thus, the ground state $\ket{\psi_0^3}$ remains inaccessible if one starts from one of the other ground states. Another interesting observation that follows from Table~{\ref{tab:prob_ta_state1}} is that the sampling probabilities of the ground states $\ket{\psi_0^2}$ and $\ket{\psi_0^4}$ are non-zero, and vary with different annealing times. This can be understood as follows. In the eigenbasis of perturbation matrix Eq.~(\ref{eq:matrix_suppressed}) $\ket{\nu_i}$, the ground state $\ket{\psi_0^1}$ can be written as 
    \begin{equation}
        \ket{\psi_0^1} = \frac{1}{\sqrt{2}} \ket{\nu_1} - \frac{1}{\sqrt{2}} \ket{\nu_4},
    \end{equation}
    where $\ket{\nu_1} = 1/2 (\sqrt{2},1,0,1), \ket{\nu_2} = 1/2 (-\sqrt{2},1,0,1), \ket{\nu_3} = (0,0,1,0)$, and $\ket{\nu_4} = 1/\sqrt{2} (0,-1,0,1)$.
    From Fig.~\ref{fig:reverse_overlap}(a), showing the overlap of the state of the system with the lowest four states of the instantaneous Hamiltonian for $T_A=100$, it is clear that except for minor fluctuations, the amplitudes present in the first and fourth instantaneous eigenvectors remain constant during the annealing process. This suggests that, restricted to the ground state subspace, the state of the system at the end of the annealing is given as
    \begin{equation}
        \ket{\psi} = \frac{1}{\sqrt{2}} e^{i\phi_1} \ket{\nu_1} + \frac{1}{\sqrt{2}} e^{i\phi_4} \ket{\nu_4},
        \label{eq:state1and4}
    \end{equation}
    where $\exp(i\phi_{1(4)})$ is the phase acquired by the first (fourth) instantaneous energy eigenstate. When measuring the sampling probabilities in the eigenbasis of the  perturbation matrix, the acquired phases $\phi_i$ become irrelevant, as the eigenstates of the perturbation matrix are mutually orthogonal. However, when measured in the computational basis, as is generally the case, all but the ground state $\ket{\psi_0^3}$ of the problem Hamiltonian has a finite overlap with $\ket{\nu_1}$ and $\ket{\nu_4}$. Mathematically,
    \begin{equation}
        \braket{\psi_0^i|\psi} = \frac{1}{\sqrt{2}} e^{i\phi_1}  \braket{\psi_0^i|\nu_1} + \frac{1}{\sqrt{2}} e^{i\phi_4}  \braket{\psi_0^i|\nu_4},
    \end{equation}
    where $i=1,2,3,4$. Thus, these individual phases result in interference, which causes the sampling probabilities of the ground states to oscillate. As the ground state $\ket{\psi_0^3}$ of the problem Hamiltonian has zero overlaps with the eigenvectors $\ket{\nu_1}$, $\ket{\nu_2}$, and $\ket{\nu_4}$ of perturbation matrix, the sampling probability of the third excited state remains zero.
    
    \begin{table}
    \begin{center}
    \caption{Sampling probabilities of the degenerate ground states $\ket{\psi_0^i}$, $i=1,2,3,4$, of the problem "230", corresponding to different annealing times $T_A$, where the reverse annealing time is chosen to be same as the forward annealing time. The initial state is chosen to be $\ket{\psi_0^1}$, the reversal distance is $s_r=0.7$, and no waiting times are added.}
    \begin{tabular}{ |c|c|c|c|c|c|c| } 
    \hline
    \textbf{State} & \bm{$T_A=10$} & \bm{$T_A=50$} & \bm{$T_A=90$}   & \bm{$T_A=100$} & \bm{$T_A=1000$} \\ 
    \hline
    $\ket{\psi_0^1}$ & 0.1891 & 0.4348 & 0.9011 & 0.0296 & 0.2614 \\ 
    \hline
    $\ket{\psi_0^2}$ & 0.4053 & 0.2822 &  0.0495 & 0.4051 & 0.3693 \\ 
    \hline
    $\ket{\psi_0^3}$ & 1.47$\times10^{-4}$ &  2.73$\times10^{-5}$ & 2.87$\times10^{-6}$ &  4.92$\times10^{-6}$ & 3.21$\times10^{-8}$ \\ 
    \hline
    $\ket{\psi_0^4}$ & 0.4031 & 0.2829 & 0.0493 & 0.4852 & 0.3692 \\ 
    \hline
    Total & 0.9976 & 0.9990 & 1.000 & 1.0000 & 1.000 \\
    \hline
    \end{tabular}
    \label{tab:prob_ta_state1}
    \end{center}
    \end{table}

    \textbf{Different reversal distances:}
    We now study the effects of varying the reversal distance on the sampling probabilities of the ground state, while keeping the annealing time fixed at $T_A=1000$, waiting time $T_W=0$, and choosing $\ket{\psi_0^1}$ as the initial state. The resulting sampling probabilities of the four ground states are shown in Table~\ref{tab:prob_sr_state1}.
    \begin{table}
    \begin{center}
    \caption{Sampling probabilities of the degenerate ground states $\ket{\psi_0^i}$, $i=1,2,3,4$, of the problem "230", corresponding to different reversal distances $s_r$, with annealing time $T_A=1000$, $\ket{\psi_0^1}$ as the initial state, and without any waiting time.}
    \begin{tabular}{ |c|c|c|c|c|c| } 
     \hline
     \textbf{State} & \bm{$s_r=0.4$} & \bm{$s_r=0.5$} & \bm{$s_r=0.6$} & \bm{$s_r=0.7$} & \bm{$s_r=0.8$} \\ 
     \hline
    $\ket{\psi_0^1}$ & 0.0703 & 0.4507 & 0.8310 & 0.2614 & 0.6147 \\ 
     \hline
     $\ket{\psi_0^2}$ & 0.2874 & 0.0513 & 0.0566 & 0.3693 & 0.1926\\ 
     \hline
     $\ket{\psi_0^3}$ & 0.0012 & 0.0010 & 1.84$\times 10^{-6}$ & 3.21$\times 10^{-8}$ & 8.66$\times 10^{-10}$ \\ 
     \hline
     $\ket{\psi_0^4}$ & 0.2739 & 0.0789 & 0.0579 & 0.3692 & 0.1926\\ 
     \hline
     Total & 0.6396 & 0.5820 & 0.9454 & 1.0000 & 1.0000\\ 
     \hline
    \end{tabular}
    \label{tab:prob_sr_state1}
    \end{center}
    \end{table}
    In this case, we note that the total success probability decreases for smaller values of reversal distances. This can be understood in relation to Fig.~\ref{fig:fairsampling_spec} for the energy spectrum of the instantaneous Hamiltonian for this problem, which shows that the positions of the anticrossings from where the state of the system can leak out of the ground state subspace. To understand this more clearly, we show the overlap of the state of the system with the lowest four eigenstates of the instantaneous Hamiltonian for $s_r=0.6$ and $s_r=0.4$ in Figs.~\ref{fig:reverse_overlap}(b) and (c), respectively.
 
    Since we choose the ground state $\ket{\psi_0^1}$ of the problem Hamiltonian as the initial state, state 1 and state 4 have an equal amplitude at the start of the annealing, in accordance with Eq.~(\ref{eq:state1and4}). While the amplitude present in the first state stays more or less constant, we see that most of the amplitude present in the fourth state gets transferred to the fifth state slightly before $s=0.6$ due to the anticrossing at $s \approx 0.62$ between the third and the fourth excited state of the instantaneous Hamiltonian (see Fig.~\ref{fig:fairsampling_spec}). From the fifth state, some of the amplitude is transferred to the sixth state. However, soon after this point the forward part of the protocol starts, and most of the amplitude is transferred back to the fourth state. Thus the final state at the end of the algorithm mainly consists of the first and the fourth states with comparable amplitudes, as was the case for the initial state.
    
    The overlap of the state with the low-lying instantaneous energy eigenstates for $s_r=0.4$ case, shown in Fig.~\ref{fig:reverse_overlap}(c) looks starkly different. In this case, we note the involvement of several higher excited states compared to that for $s_r=0.6$. This can once again be understood on the basis of the energy spectrum of this problem (Fig.~\ref{fig:fairsampling_spec}). As before, the system starts in an initial state which is an equal superposition of the first and fourth instantaneous eigenstates. Following the respective anticrossings between the energy levels, the amplitude present in the fourth state gets sequentially transferred to the higher excited states, although Fig.~\ref{fig:fairsampling_spec} only shows up till the tenth energy level. On the other hand, part of the amplitude present in the first state is shifted to the second state at the anticrossing between these two levels at $s \approx 0.42$. While most of the transferred amplitude from the first state returns to the first state in the forward part of the anneal, the final amplitude in the fourth stays small.
    
    As in the case of varying annealing times, we find that the final sampling probabilities, which are measured in the computational basis, fluctuate, except for the ground state $\ket{\psi_0^3}$ whose sampling probability stays fairly low for all values of reversal distances. This can be explained based on the interference of the accumulated phases in the state of the system. Furthermore, from Table~\ref{tab:prob_sr_state1} we note that the sampling probability of the ground state $\ket{\psi_0^3}$ increases as the value of the reversal distance is lowered. This is due to the fact that the third state of the instantaneous Hamiltonian becomes accessible via the higher excited states or the anticrossings within the ground state subspace as the $s$ values are made small.
    
    \begin{figure*}
    \begin{minipage}{0.49\textwidth}
         \centering
         \includegraphics[width=\textwidth]{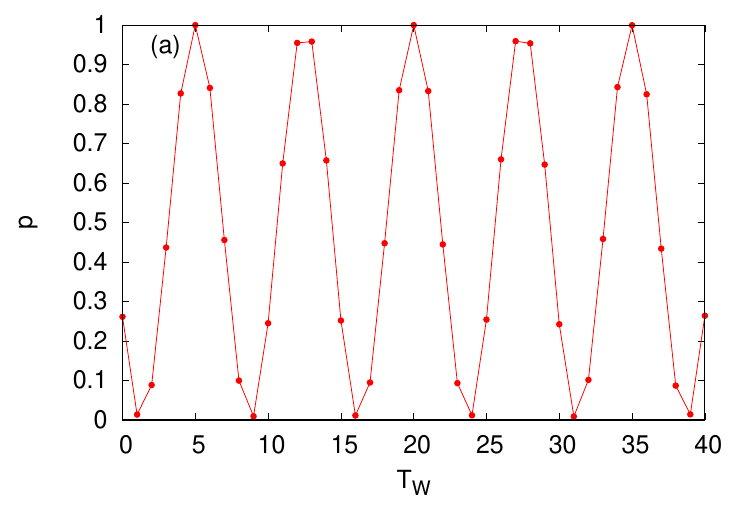}
     \end{minipage}
     \hfill
     \begin{minipage}{0.49\textwidth}
         \centering
         \includegraphics[width=\textwidth]{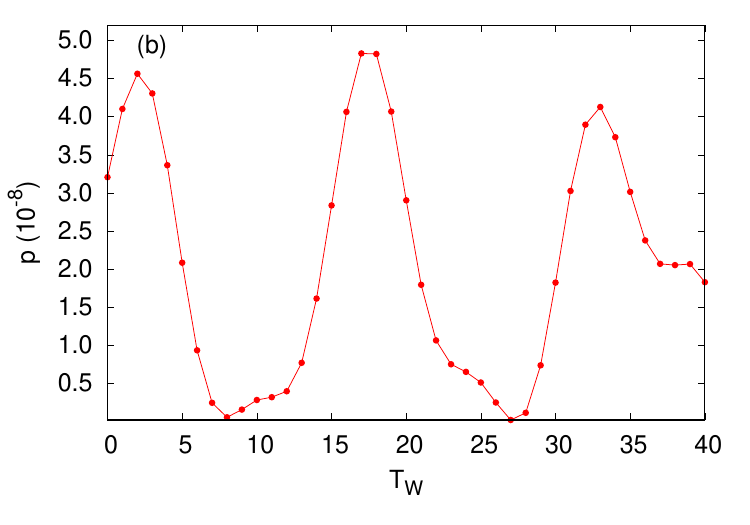}
     \end{minipage}
     \caption{(Color online) Success probability of (a) the ground state $\ket{\psi_0^1}$ and (b) $\ket{\psi_0^3}$ as a function of different waiting times for $T_A=1000$.}
    \label{fig:fairsampling_wait}
    \end{figure*}

    \textbf{Different waiting times:}
    After having studied the effects of varying the annealing times and the reversal distances on the sampling probabilities of the reverse annealing algorithm, we now perform a similar analysis, but by altering the waiting times $T_W$ from 0 to 40. For this case, we choose annealing time $T_A=1000$, reversal distance $s_r=0.7$, and the ground state $\ket{\psi_0^1}$ as the initial state. The resulting behavior of the sampling probabilities of the ground states $\ket{\psi_0^1}$ and $\ket{\psi_0^3}$ is shown in Fig.~\ref{fig:fairsampling_wait}. Although on very different magnitudes, we observe oscillations in the sampling probabilities of both these states. This, once again, makes apparent that the accumulation of different phases in the amplitudes of the wavefunction (expressed in the instantaneous energy eigenbasis) leads to interference, resulting in fluctuations in the sampling probabilities. For the chosen set of parameters, we find that the total success probability always remains close to one. Furthermore, since the ground state $\ket{\psi_0^3}$ is decoupled from the rest of the ground state subspace according to perturbation matrix Eq.~(\ref{eq:matrix_suppressed}), the sampling probability of $\ket{\psi_0^3}$ stays on the order of $\mathcal{O}(10^{-8})$.

    \textbf{A different initial state:} 
    So far, we have only focused on the results for which the ground state $\ket{\psi_0^1}$ was chosen as the initial state. We now discuss the sampling results obtained from the reverse annealing protocol starting from the ground state $\ket{\psi_0^3}$. Keeping the reversal distance fixed as $s_r=0.7$, and waiting time to zero, we study the effect of varying the annealing time on the sampling probabilities. The resulting sampling probabilities are shown in Table~\ref{tab:prob_ta_state3}. Unlike the case shown in Table~\ref{tab:prob_ta_state1} where $\ket{\psi_0^1}$ is chosen as the initial state, choosing $\ket{\psi_0^3}$ as the initial state does not cause the sampling probability of the initial state to redistribute to the other ground states, i.e., $\ket{\psi_0^3}$ is sampled with a probability close to 1. This can once again be understood on the basis of Eq.~(\ref{eq:matrix_suppressed}), from where it is evident that close to $s \approx 1$, the instantaneous second excited state, which corresponds to the ground state $\ket{\psi_0^3}$ of the problem Hamiltonian, is completely decoupled from the rest of the four low-lying instantaneous energy eigenstates. Therefore, the amplitude present in the ground state $\ket{\psi_0^3}$ of the problem Hamiltonian cannot be transferred to the rest of the four lowest-lying energy states.
    
    \begin{table}
    \begin{center}
    \caption{Sampling probabilities of the degenerate ground states $\ket{\psi_0^i}$, $i=1,2,3,4$, of the problem "230", corresponding to different annealing times $T_A$, where the reverse annealing time is chosen to be same as the forward annealing time. The initial state is chosen to be $\ket{\psi_0^3}$, the reversal distance is $s_r=0.7$, and no waiting times are added.}
    \begin{tabular}{ |c|c|c|c| } 
     \hline
     \textbf{State} & \bm{$T_A=10$} & \bm{$T_A=100$} & \bm{$T_A=1000$}\\ 
     \hline
     $\ket{\psi_0^1}$ & 1.47$\times 10^{-4}$ & 4.92$\times 10^{-6}$ & 3.21$\times 10^{-8}$ \\ 
     \hline
     $\ket{\psi_0^2}$ & 3.07$\times 10^{-4}$ & 3.71$\times 10^{-3}$ & 1.97$\times 10^{-5}$\\ 
     \hline
     $\ket{\psi_0^3}$ & 0.9963 & 0.9928 & 0.9996 \\ 
     \hline
     $\ket{\psi_0^4}$ & 4.28$\times 10^{-4}$ & 3.41$\times 10^{-3}$ & 1.72$\times 10^{-5}$\\ 
     \hline
     Total & 0.9968 & 1.0000 & 1.0000\\ 
     \hline
    \end{tabular}
    \label{tab:prob_ta_state3}
    \end{center}
    \end{table}
    
    We note similar observations for the case where the reversal distance is varied keeping $T_A=1000$ and $T_W=0$, and the ground state $\ket{\psi_0^3}$ is chosen as the initial state. In this case, the state $\ket{\psi_0^3}$ is sampled with a probability close to one for large values of the reversal distances. However, upon lowering the value of $s_r$ the total success probability of the ground state decreases, and the other ground states can become accessible via the anticrossings leading to the higher excited states.

\subsubsection{D-Wave results}
After having discussed the sampling behavior of the ideal implementation of reverse annealing protocol, in this section, we discuss the corresponding results obtained from DWAdv. As before, we discuss the effects of varying the different annealing controls available in the D-Wave systems on the sampling probabilities of the 14-variable problem "230".

    \textbf{Different annealing times:} Keeping the reversal distance fixed at $s_r=0.7$ and waiting time at $T_W=0$, and choosing the ground state $\ket{\psi_0^1}$ of the problem Hamiltonian $\ket{\psi_0^1}$ as the initial state, we start by studying the sampling efficiency of the reverse annealing protocol on DWAdv by varying the annealing times. As before, we set the total number of samples to 1000. The resulting sampling probabilities are given in Table~\ref{tab:prob_ta_state1_dw}. The first observation that follows is that the total success probability stays close to one for all annealing times. Moreover, unlike the case for standard quantum annealing, the sampling probabilities obtained with DWAdv using reverse annealing are \textbf{not fair}. For all the values of the annealing time chosen, we find that the sampling probability of the ground state $\ket{\psi_0^3}$ is totally suppressed. Such a behavior resembles the results obtained from both perturbation theory as well as the simulation results, suggesting that even if the non-ideal elements like noise and temperature effects are present in this regime, the state $\ket{\psi_0^3}$ is inaccessible in the quantum annealer for $s \geq 0.6$ when starting from $\ket{\psi_0^1}$. 
    %In addition, as was the case for the corresponding simulation results, we find that despite much longer annealing times on DWAdv compared to those in the simulation, the sampling probabilities of the ground states $\ket{\psi_0^2}$ and $\ket{\psi_0^4}$ of the problem Hamiltonian are non-zero and can fluctuate.
   However, unlike the case of simulations where the sampling probabilities of ground states $\ket{\psi_0^1}$, $\ket{\psi_0^2}$ and $\ket{\psi_0^3}$ were fluctuating with the annealing time, we note that the sampling probability of ground state $\ket{\psi_0^1}$ obtained from DWAdv is decreasing while those of states $\ket{\psi_0^2}$ and $\ket{\psi_0^4}$ are increasing with an increasing annealing time. Understanding the cause of such a behavior calls for further investigation.
    
    \begin{table}[!htp]
    \begin{center}
    \caption{Sampling probabilities of the degenerate ground states $\ket{\psi_0^i}$, $i=1,2,3,4$, of the problem "230" obtained using DWAdv, corresponding to different annealing times $T_A$, where the reverse annealing time is chosen to be same as the forward annealing time. The initial state is chosen to be $\ket{\psi_0^1}$, the reversal distance is $s_r=0.7$, and no waiting times are added.}
    \resizebox{\columnwidth}{!}{
    \begin{tabular}{ |c|c|c|c|c|c|c| } 
     \hline
     \textbf{State} & \bm{$T_A=0.5\mu s$} & \bm{$T_A=10\mu s$} & \bm{$T_A=50\mu s$} & \bm{$T_A=200\mu s$} \\ 
     \hline
     $\ket{\psi_0^1}$ & 0.9991 & 0.9805 & 0.6709 & 0.5405 \\ 
     \hline
     $\ket{\psi_0^2}$ & 0.0006 & 0.0091 & 0.1470 & 0.2490 \\ 
     \hline
     $\ket{\psi_0^3}$ & 0 & 0 & 0 & 0 \\ 
     \hline
     $\ket{\psi_0^4}$ & 0.0003 & 0.0099 & 0.0830 & 0.2105 \\ 
     \hline
     Total & 1.0000 & 0.9995 & 0.9909 & 1.0000 \\ 
     \hline
    \end{tabular}
    }
    \label{tab:prob_ta_state1_dw}
    \end{center}
    \end{table}
    
    \textbf{Different reversal distances:} Moving next to the effects of varying the reversal distance on the sampling behavior of the reverse annealing protocol on DWAdv, we set the reverse and forward annealing times to $T_A=20~\mu s$ and $T_W=0$, and as before, start the protocol with the ground state $\ket{\psi_0^1}$ of the problem Hamiltonian. The corresponding results are shown in Table~\ref{tab:prob_sr_state1_dw}, from where we note that the total success probability is close to one for large values of the reversal distance and decreases for the smaller values of the reversal distance. Although the annealing scheme implemented by the D-Wave annealers is not linear, the energy spectrum of the problem obtained using the linear scheme (see Fig.~\ref{fig:fairsampling_spec}) seems to capture the main trend of the total success probability well. As the value of the reversal distance is decreased, the chances that the amplitude present in the ground state subspace of the instantaneous Hamiltonian leaks to the higher excited states increases. Another important observation from Table~\ref{tab:prob_ta_state1_dw} is that the sampling probability of the ground state $\ket{\psi_0^3}$ stays small, especially for large values of the reversal distance. Nevertheless, as the value of the reversal distance is lowered, $\ket{\psi_0^3}$ is noted to have an increasingly large sampling probability. Such a behavior was also observed from the corresponding ideal simulations. %Such behavior is expected from ideal quantum annealing since none of the other ground states have an overlap with the ground state $\ket{\psi_0^3}$.  can become accessible via the higher excited states which can have a non-zero overlap with it, as explained above. Additionally, due to the interference in the acquired phases in the different components of the wavevector, the sampling probabilities of the other three ground state can fluctuate, as is observed to be the case. 
    
    \begin{table}[!htp]
    \begin{center}
    \caption{Sampling probabilities of the degenerate ground states $\ket{\psi_0^i}$, $i=1,2,3,4$, of the problem "230" obtained using DWAdv, corresponding to different reversal distances $s_r$, with annealing time $T_A=20\mu s$, $\ket{\psi_0^1}$ as the initial state, and without any waiting time.}
    \resizebox{\columnwidth}{!}{
    \begin{tabular}{ |c|c|c|c|c|c|c| }
     \hline
     \textbf{State} & \bm{$s_r=0.3$} & \bm{$s_r=0.4$} & \bm{$s_r=0.5$} & \bm{$s_r=0.6$} & \bm{$s_r=0.7$} & \bm{$s_r=0.8$} \\ 
     \hline
    $\ket{\psi_0^1}$ & 0.2173 & 0.4080 & 0.3412 & 0.3120 & 0.9287 & 1.0000\\ 
     \hline
     $\ket{\psi_0^2}$ & 0.2468 & 0.3235 & 0.4269 & 0.4075 & 0.0401 & 0\\ 
     \hline
     $\ket{\psi_0^3}$ & 0.1725 & 0.0065 & 0.0005 & 0 & 0 & 0 \\ 
     \hline
     $\ket{\psi_0^4}$ & 0.2665 & 0.2556 & 0.2279 & 0.2751 & 0.0295 & 0\\ 
     \hline
     Total & 0.9031 & 0.9936 & 0.9965 & 0.9946 & 0.9983 & 1.0000\\ 
     \hline
    \end{tabular}
    }
    \label{tab:prob_sr_state1_dw}
    \end{center}
    \end{table}

    \textbf{Different waiting times:} Next, we study the effects of varying the waiting time on the sampling probabilities of the ground state of the degenerate problem under observation. For this, we choose $T_A=20~\mu s$, $s_r=0.7$, and $\ket{\psi_0^1}$ as the initial state. As for the case of varying $T_A$, we find that the total success probability for all values of $T_W$ is close to one, due to the value of the reversal distance being large. Similar to the results described above, we find that the sampling probability  of $\ket{\psi_0^3}$ stays zero.

    \textbf{A different initial state:} Lastly, we discuss the case of varying the annealing controls when the protocol starts with the ground state $\ket{\psi_0^3}$ of the problem Hamiltonian $\ket{\psi_0^3}$. As for the corresponding results obtained numerically, in this case, we find that for a large value of the reversal distance the ground state $\ket{\psi_0^3}$ is sampled with a probability close to one. %This is in agreement with perturbation theory, according to which starting from this state it is not possible to access the other ground states in the ground state subspace (see Eq.~(\ref{eq:matrix_suppressed})). However, when $s_r$ values are lowered, the state of the system can leak out of the ground state subspace, so that the total success probability is not one, but the other ground states can be sampled due to their accessibility via the higher excited states.

% \section{Scaling performance}

% \section{Sampling efficiency}
% \label{sec:sampling}

% \subsection{Reverse annealing}

\section{Conclusion}
\label{sec:conclusion}

There are various metrics using which the performance of a heuristic approach for solving an optimization problem can be gauged. The focus of this paper was to assess the performance of quantum annealing in solving problems with more than one feasible solution. To this end, we used both numerical and physical implementation of standard as well as reverse annealing protocols, as offered by the D-Wave quantum annealers to solve a set of specially designed 2-SAT problems with four known solutions. We then used the scaling of time to solution (TTS) and the efficiency of the method to sample the four solutions as the relevant measures for the performance. It is worth mentioning that although 2-SAT problems are not NP-hard, finding all the solutions of a 2-SAT problem with multiple solutions is.

%For optimization problems with an unknown ground state, the efficiency of a method can be analyzed in terms of the lowest energy value that it samples. On the other hand, if the ground state of an optimization problem is known, a good performance metric is the probability with which the ground state is obtained. Another related quantifier is the time to solution (TTS) which is the run-time required to obtain the solution of the optimization problem at least once with a certain probability. An important aspect for assessing the suitability of a certain approach is to study the scaling of its metrics like success probability or TTS as the size of the problems grows. Furthermore, for problems that have more than one solution, another relevant performance metric is the efficiency with which a certain heuristic method samples all the solutions. Using a newly constructed set of 2-SAT problems with four satisfying assignments and up to $N=20$ variables, we studied the scaling behavior and sampling efficiency of quantum annealing for these problems. To this end, we used both simulations as well as the D-Wave quantum annealer DWAdv. Furthermore, in addition to the standard quantum annealing (as implemented by D-Wave), we employed reverse annealing in our analysis using both the approaches. In doing so, we implemented similar controls in our simulations for the reverse annealing protocol as those available on DWAdv. These include a choice for the annealing time, reversal distance, waiting time, and the initial state. 

Restricting ourselves first to the standard quantum annealing algorithm, we found that the sampling probabilities of the four ground states were in agreement with the predictions from perturbation theory if the chosen annealing time was sufficiently long. From this observation, the sampling probabilities could be expected to be more fair with the inclusion of the higher-order coupling terms in the initial Hamiltonian. On the other hand, despite of choosing much longer annealing times, the sampling probabilities resulting sampling probabilities from DWAdv were roughly uniform and therefore different from our results from the simulations as well as the perturbative analysis. Although advantageous in this case, such a deviation in the sampling behavior hints towards the presence of certain non-ideal mechanisms during the evolution of the state of the system in the annealer.

Regarding the scaling aspect, using standard quantum annealing for these problems, we observed an exponentially growing TTS99 with increasing size of the problems, for our ideal simulations as well as for the quantum annealer. However, although the scaling exponent for the former was found to be worse compared to even a brute force search for the ground state in the long annealing time limit, the scaling exponent obtained from DWAdv was significantly smaller. Furthermore, it was found that the scaling behavior from the annealer could fit well to equilibrium probability distribution using $\beta = 1/(k_B T)$ as the fitting parameter.

The sampling results using the reverse annealing protocol obtained with the simulations and DWAdv were found to be in close agreement, and the sampling probabilities depended greatly on the values of the annealing controls. While the sampling probabilities resulting from the simulations could once again be justified on the basis of perturbation theory when the chosen annealing times were sufficiently long, understanding the mechanisms leading to seemingly similar behavior from the D-Wave quantum annealer calls for a careful deeper investigation. 

%These observations could be justified fundamentally making use of perturbation theory and the numerically obtained energy spectrum of the problems. The similarity between the numerical results and those obtained using DWAdv suggests that in this regime, the noise and temperature effects play a less dominant role in the evolution of the state of the system in the DWAdv.

\section{Acknowledgements}
The authors gratefully acknowledge the Gauss Centre for Supercomputing e.V. (www.gauss-centre.eu) for funding this project by providing computing time through the John von Neumann Institute for Computing (NIC) on the GCS Supercomputer JUWELS \cite{JUWELS} at Jülich Supercomputing Centre (JSC). 
% The authors also gratefully acknowledge the computing time granted through JARA on the supercomputer JURECA \cite{jureca} at Forschungszentrum Jülich. 
V.M. acknowledges support from the project JUNIQ funded by the German Federal Ministry of Education and Research (BMBF) and the Ministry of Culture and Science of the State of North Rhine-Westphalia (MKW-NRW) and from the project EPIQ funded by MKW-NRW.

\appendix
\section{2-SAT problem instances}
\label{sec:appendix_problems}
In this appendix we list the three 14-variable problem instances that have been discussed in this work as having a fair sampling (problem "1"), having an unfair sampling (problem "3"), and a problem with zero theoretical sampling probability of one of the ground states (problem "230"). The clauses constituting the three SAT problems are given in Table~\ref{tab:SATproblems}.
\begin{table}[!htp]
    \centering
    \caption{Three instances of 2-SAT problems: Problem "1" with almost fair sampling, Problem "3" with unequal sampling probabilities of the four ground states, and Problem "230" with zero sampling probability of one of the ground states.}
    \begin{tabular}{|p{1cm}||p{1cm}|p{1cm}||p{1cm}|p{1cm}||p{1cm}|p{1cm}|}
    \hline
    {Clause} & \multicolumn{2}{|c||}{Problem:"1"} &\multicolumn{2}{c||}{Problem:"3"} & \multicolumn{2}{c|}{ Problem:"230"}\\
    \hline 
    1 & $\overline{x_{13}}$ & $x_{14}$ & $\overline{x_8}$ & $x_{10}$ & $\overline{x_{10}}$ & $\overline{x_{13}}$ \\ 
    \hline 
    2 & $\overline{x_{11}}$ & $x_{13}$ & $\overline{x_7}$ & $x_{14}$ & $\overline{x_8}$  & $x_{13}$ \\ 
    \hline 
    3 & $\overline{x_{10}}$ & $x_{12}$ & $\overline{x_6}$ & $\overline{x_{11}}$ & $\overline{x_7}$ & $\overline{x_{14}}$ \\
    \hline 
    4 & $\overline{x_6}$ & $\overline{x_8}$ & $\overline{x_4}$ & $x_{11}$ & $\overline{x_6}$ & $\overline{x_{14}}$ \\
    \hline 
    5 & $\overline{x_6}$ & $x_{11}$ & $\overline{x_3}$ & $\overline{x_5}$ & $\overline{x_5}$ & $\overline{x_{12}}$ \\ 
    \hline 
    6 & $\overline{x_4}$ & $\overline{x_6}$ & $\overline{x_2}$ & $x_{14}$ & $\overline{x_4}$ & $\overline{x_{8}}$ \\
    \hline 
    7 & $\overline{x_4}$ & $x_{10}$ & $\overline{x_1}$ & $\overline{x_{11}}$ & $\overline{x_2}$ & $x_9$ \\ 
    \hline 
    8 & $\overline{x_2}$ & $x_{14}$ & $\overline{x_1}$ & $x_4$ & $\overline{x_2}$ & $x_{12}$ \\ 
    \hline 
    9 & $\overline{x_1}$ & $x_{5}$ & $x_2$ & $\overline{x_{12}}$ & $\overline{x_1}$ & $\overline{x_{11}}$ \\ 
    \hline 
    10 & $x_1$ & $x_{9}$ & $x_3$ & $x_6$ & $x_1$ & $\overline{x_3}$ \\ 
    \hline 
    11 & $x_3$ & $\overline{x_5}$ & $x_4$ & $\overline{x_{14}}$ & $x_2$ & $x_8$ \\ 
    \hline 
    12 & $x_4$ & $\overline{x_{14}}$ & $x_6$ & $\overline{x_{11}}$ & $x_3$ & $\overline{x_{12}}$ \\
    \hline 
    13 & $x_4$ & $\overline{x_9}$ & $x_7$ & $\overline{x_{10}}$ & $x_4$ & $\overline{x_9}$ \\ 
    \hline 
    14 & $x_6$ & $\overline{x_{10}}$ & $x_8$ & $\overline{x_{13}}$ & $x_6$ & $x_{11}$ \\
    \hline 
    15 & $x_7$ & $x_{13}$ & $x_8$ & $\overline{x_9}$ & $x_{12}$ & $\overline{x_{13}}$ \\ 
    \hline 
    \end{tabular} 
    \label{tab:SATproblems}
\end{table}

\section{Transition probability as a function of annealing time}
\label{app:trans_prob}

\begin{figure}
\centering
\includegraphics[scale=0.7]{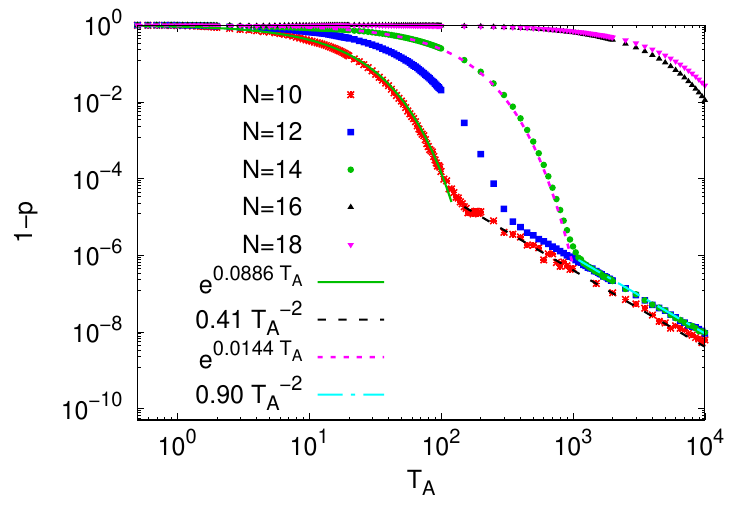}
     \caption{(Color online) Transition probability $1-p$ as a function of annealing time for chosen instances of problems with different $N$.}
\label{fig:tscan}
\end{figure}

To understand the scaling behavior of the TTS99 from the simulations \ref{sec:scaling_sim}, in Fig.~\ref{fig:tscan} we show the transition probability $1-p$ as a function of the annealing time for instances of $N=10$ and $N=14$-variable 2-SAT problems that constitute the median success probability of the sets. From the figure, it is evident that both problems show two distinct behaviors of the transition probability as the annealing time increases. At first, the transition probability decreases exponentially with $T_A$ for small values of $T_A$, however, for longer times it exhibits a polynomial ($\mathcal{O}(T_A^{-2})$) dependence. It is remarkable that in spite of the fact that, unlike in a simple two level-system, the ground state of the problem Hamiltonian is four-fold degenerate, this behavior is very similar to the one observed for a two-level system \cite{Nishimori2008}. The first region reflects the Landau-Zener transition \cite{landau1932theorie, zener1932non, de1997theory}, while in the second region the transition probability decreases as $1/T^2_A$, as expected from the adiabatic theorem \cite{teufel2003adiabatic, jansen2007bounds, Lidar_albash2018,Mehta:2023cbh}. Moreover, from Fig.~\ref{fig:tscan} we see that the point transition to the adiabatic region shifts to larger values of annealing times as the size of the problems increases. Making use of this information we can make a few remarks about the behavior of TTS99 observed in Fig.~\ref{fig:tts99_deg_simulations}. In the second region, i.e., for small $N$ and large $T_A$ values, the R.H.S. of Eq.~(\ref{eq:TTS}) is proportional to $T_A/(\ln C - 2\ln T_A))\approx -T_A/2\ln T_A$ because $T_A\gg C$. Here $C$ is a parameter that depends on $N$. For small problem sizes $N$, TTS99 remains constant for a fixed value of $T_A$ (e.g., $T_A=1000$), see Fig.~\ref{fig:tts99_deg_simulations}. For the larger problems, where both $T_A=100, 1000$ lie in the first region ($1-p = \exp(-C' T_A)$ where $C'$ is a fitting parameter ) as seen for the $N=14$ points in Fig.~\ref{fig:tscan}, the R.H.S. of Eq.~(\ref{eq:TTS}) becomes proportional to $1/C'$. After this point, the scaling of TTS99 depends on the scaling of $C'$, which in turn depends on the intricate properties of the energy spectrum. This suggests that in this region the TTS values for $T_A=100,1000$ should coincide, in concert with Fig.~\ref{fig:tts99_deg_simulations}. Moreover, as also noted from \cite{paper1,paper2} $T_A=10$ corresponds to a special case for these 2-SAT problems where various non-adiabatic mechanisms play a prominent role during the evolution that enhance the success probability, and thus the picture described above does not necessarily hold for this case. %In addition, it is worth mentioning that for most real-world applications, choosing annealing times that yield $1-p \approx 10^{-2}$ is more than sufficient for most practical purposes.

\section{Scaling results for other problems}
\label{app:otherprobs}
As seen in section~\ref{sec:scaling_dwave}, the scaling of TTS99 obtained from DWAdv for the 2-SAT problems shows a good agreement with the analytical expression for the equilibrium distribution given by Eq.~(\ref{eq:equil}). To further increase our confidence in this conjecture, we extend our analysis to two other sets of problems: one derived from the original set of 2-SAT problems, and the ferromagnetic spin chain problem, as described below.

\subsection{Rescaled 2-SAT problems}

\begin{figure}
\centering
\includegraphics[scale=0.7]{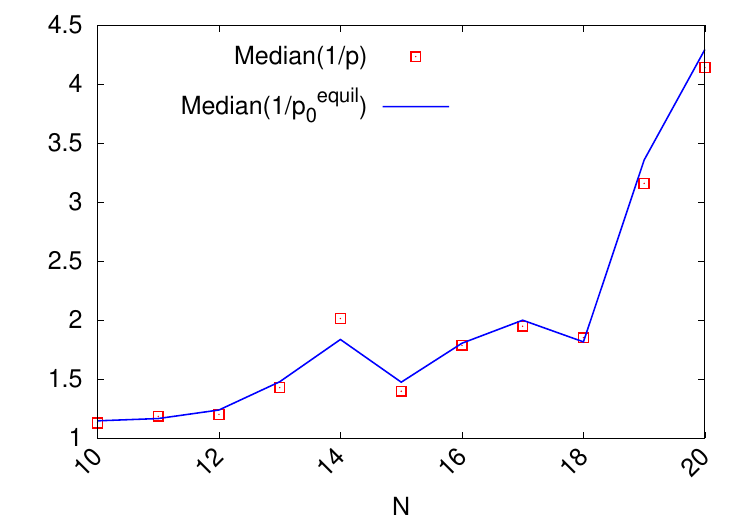}
\caption{(Color online) Comparison of scaling of the mean and the median cases obtained from DWAdv with those of the equilibrium 1/p for $\alpha =0.25$ with $T_A=1000$~$\mu s$.}
\label{fig:alpha_0.25}
\end{figure}

\begin{figure}
\centering
\includegraphics[scale=0.7]{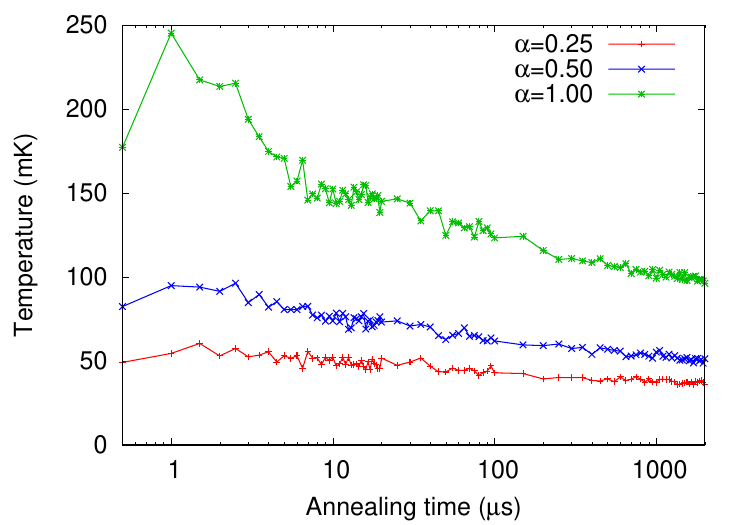}
\caption{(Color online) Effective temperature calculated using the probabilities resulting from DWAdv for one of the 20-variable 2-SAT problems found in the median for the $\alpha=1$ set. The same problem was used to perform the same analysis for $\alpha=0.25$ and $\alpha=0.50$.}
\label{fig:temp_comb}
\end{figure}

We continue the investigation by introducing a parameter $\alpha$ for rescaling the problem Hamiltonian $H_P$ corresponding to the 2-SAT problems. Choosing $0 < \alpha < 1$ reduces the energy gap $\Delta E$ between the lowest two levels of the 2-SAT problem. Setting $\alpha=0.25$ and selecting the median cases corresponding to the original set, Fig.~\ref{fig:alpha_0.25} shows the resulting comparison of the inverse success probabilities $1/p$ from DWAdv and $1/p_0^{equil}$, corresponding to a much longer annealing time $T_A=1000$~$\mu s$. We note a better agreement between the two values in this case, suggesting that the sampling probabilities from the D-Wave annealer approximately match those from the equilibrium distribution. For this case, we obtain $\beta = 5.52$ which corresponds to a temperature of about $37$~mK, which is of a similar order as the physical temperature of about $12$~mK, typical for the DWAdv annealer. To understand the reasons for the differences in the values of the $\beta$ parameter and, consequently, the temperature values for different values of the rescaling factor $\alpha$, in Fig.~\ref{fig:temp_comb} we show the fit temperatures for various annealing times. The plot suggests a slower convergence to equilibrium with increasing values of $\alpha$, i.e., for larger gaps $\Delta E$ between the lowest two levels of the 2-SAT problems. Furthermore, since for $\alpha = 1$, we choose a relatively short annealing time $T_A=20$~$\mu$s, the system is far from attaining equilibrium in this case, and therefore, the corresponding measure of temperature is significantly different from the system temperature.

\subsection{Ferromagnetic spin chain problem}
\begin{figure}
\centering
\includegraphics[scale=0.7]{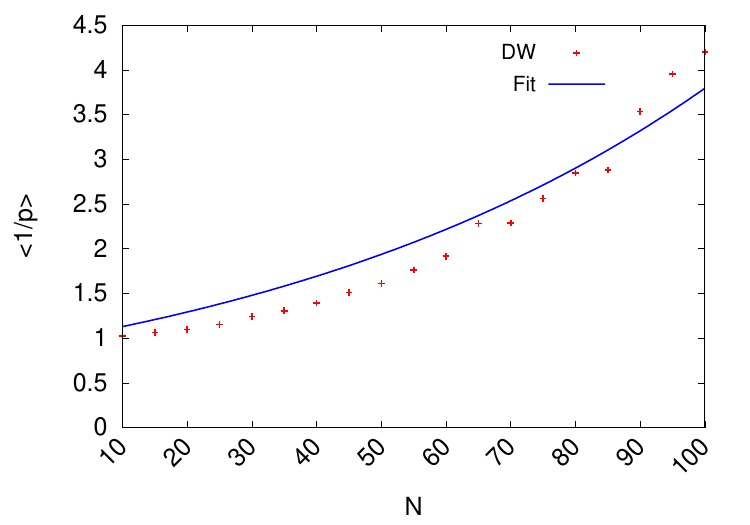}
\caption{(Color online) Comparison of scaling of $\langle1/p\rangle$ obtained from DWAdv against the fitting function $\sum_{n=0}^5 {N-1 \choose n}\exp(-n\beta \Delta E)$ as a function of system size $N$ for the ferromagnetic spin chain problem with $T_A=500~\mu$s.}
\label{fig:ferrochain}
\end{figure}

In our analysis so far, we have found the scaling of TTS99 from the D-Wave annealers to be exponential for our 2-SAT problems, and we argued that this behavior is related to the exponentially increasing degeneracy of the first excited state of these problems with their increasing size. To further test our hypothesis, we create simple instances of spins connected via ferromagnetic couplings with various sizes, that have an increasing first excited state degeneracy. Each problem consists of $N$ spins that are connected via a ferromagnetic coupling $J=-0.5$, with $10 \leq N \leq 100$. The ground state of these problems is two-fold degenerate and the ground state energy is $J(N-1)$. The energy gap between the ground state and the $n$th excited states is $2nJ$ and the degeneracy of the $n$th excited state is $2{N-1 \choose n}$. Furthermore, for improved statistics, we create 200 spin-reversal instances for each $N$ by randomly selecting a few spins and flipping the sign of the coupling between the chosen spin and its neighbors. Doing this only alters the ground state of the new problem, while preserving the other properties of the original problem. Figure~\ref{fig:ferrochain} shows the results for the analytical expression for $\langle1/p\rangle$ including also the second excited states in the partition function fit the mean inverse success probabilities obtained from DWAdv well.

\bibliography{references}
\end{document}